\documentclass[useAMS,usegraphicx,usenatbib,referee]{mn2e}

\title[Non-linear particle acceleration at shock waves]
{A general solution to non-linear 
particle acceleration at non-relativistic shock waves}
\author[E. Amato and P. Blasi]{E. Amato$^{1}$\thanks{E-mail:
amato@arcetri.astro.it} 
and P. Blasi$^{1}$\thanks{E-mail: blasi@arcetri.astro.it}\\
$^{1}$INAF-Osservatorio Astrofisico di Arcetri, 
Largo E. Fermi, 5, 50125, Firenze, Italy}

\begin{document}

\date{Accepted ----. Received -----}

%\pagerange{\pageref{firstpage}--\pageref{lastpage}} \pubyear{2005}

\maketitle

\label{firstpage}

\begin{abstract}
Diffusive acceleration at collisionless shock waves remains one of the 
most promising acceleration mechanisms for the description of the 
origin of cosmic rays at all energies. A crucial ingredient to be
taken into account is the reaction of accelerated particles on the
shock, which in turn determines the efficiency of the process. Here
we propose a semi-analytical kinetic method that allows us to calculate
the shock modification induced by accelerated particles together with 
the efficiency for particle acceleration and the spectra of
accelerated particles. The shock modification is calculated
for arbitrary environment parameters (Mach number, maximum momentum, 
density) and for arbitrary diffusion properties of the medium.
Several dependences of the diffusion coefficient on particle momentum and
location are considered to assess the goodness of the method. 
\end{abstract}

\begin{keywords}
acceleration of particles - shock waves
\end{keywords}

\section{Introduction}
Most scenarios for the origin of cosmic rays rely upon the 
acceleration of charged particles in the presence of shock 
waves, developed in sources such as supernova remnants, active
galaxies, planetary shocks, gamma ray bursts and many others. 
The basic features of the acceleration process have been highlighted
in the pioneering papers by \cite{krymskii,bo78,bell78} in the
context of the so-called {\it test particle} assumption. Several
excellent reviews (\cite{drury83,be87,je91,maldru2001}) 
discuss in detail the many problems that are still open concerning
the acceleration process. Among these, a fundamental one is the
limited applicability of the results found within the test particles
approach. In most scenarios for the origin of either galactic
or extra-galactic cosmic rays, in fact, an appreciable fraction of the 
kinetic energy crossing the shock needs to be transfered to 
accelerated particles. This need contradicts the very assumption
that the accelerated particles are test particles, unable to 
exert any dynamical reaction onto the shocked fluid. The well
known result that the spectrum of the accelerated particles is
a power law with slope nearly independent of the detailed properties of 
the system (e.g. diffusion coefficient) holds only within the
context of this test particle approximation. Relaxing this 
assumption leads to the modification of the shock by the accelerated
particles, a phenomenon that has received much attention in
the context of the so-called two-fluid models (\cite{dr_v80,dr_v81}), 
kinetic models (\cite{malkov1,malkov2,blasi1,blasi2}) and numerical
approaches, both Monte Carlo and other simulation procedures 
(\cite{je91,bell87,elli90,ebj95,ebj96,kj97,kj05,jones02}).
For an accurate recent review see the work by \cite{maldru2001},
from which the reader can see the weak and strong points of 
each approach. The present paper illustrates a kinetic 
analytical approach, which provides the exact solution for 
the spectrum of accelerated particles and shock modification 
in a very general situation in which the diffusion properties
of the medium are arbitrary.
The problem is reduced to solving an integral-differential 
equation, which easily leads to the required solution. For the 
injection of particles at the shock surface we implement
the recipe previously presented by \cite{vannoni}. In all
the cases that we considered we never find evidence for multiple
solutions. The method we propose is of general validity, in that 
it can be used for an arbitrary momentum dependence
of the diffusion coefficient and for diffusion properties (related
to the magnetization properties of the medium) that can change in 
an arbitrary way with the spatial location in the fluid. 

\section[]{Calculations}
The equations for the conservation of the mass and momentum fluxes
between upstream infinity and a point $x$ in the upstream region
can be written as:
\begin{equation}
\rho_0 u_0 = \rho(x) u(x),
\label{eq:masscons} 
\end{equation}
\begin{equation}
\rho_0 u_0^2 + P_{g,0} = \rho(x) u(x)^2 + P_g(x) + P_{CR} (x)\ ,
\label{eq:momcons}
\end{equation}
where $\rho$, $u$ and $P_g$ are the gas density, velocity and 
pressure (the corresponding quantities at upstream infinity 
are indicated with the index $0$). The pressure of accelerated
particles is defined as
\begin{equation}
P_{CR}(x) = \frac{1}{3} \int_{p_{inj}}^{p_{max}} dp\ 4 \pi p^3 v(p) f(x,p),
\end{equation}
and $f(x,p)$ is the distribution function of accelerated particles. 
Here $p_{inj}$ and $p_{max}$ are the injection and maximum momentum. The 
function $f$ vanishes at upstream infinity, which implies that there 
are no cosmic rays infinitely distant from the shock in the upstream 
region. The distribution function satisfies the following transport 
equation in the reference frame of the shock:
\begin{equation}
\frac{\partial}{\partial x}
\left[ D(x,p)  \frac{\partial}{\partial x} f(x,p) \right] - 
u  \frac{\partial f (x,p)}{\partial x} + 
\frac{1}{3} \left(\frac{d u}{d x}\right)
~p~\frac{\partial f(x,p)}{\partial p} + Q(x,p) = 0.
\label{eq:trans}
\end{equation}
The $x$ axis is oriented from upstream infinity ($x=-\infty$) to
downstream infinity ($x=+\infty$), with the shock located at $x=0$.
The injection is introduced here through the function $Q(x,p)$. The
diffusion properties are described by the arbitrary function $D(x,p)$,
depending on both momentum and space. In previous approaches
restrictive assumptions on the diffusion coefficient were always 
adopted in order to facilitate the path to the solution. 
The solution $f$ can be written in the following implicit form:
$$
f(x,p) = \exp\left[-\int_x^0 dx' \frac{u(x')}{D(x',p)}\right] \times
$$
\begin{equation}
\left\{
f_0(p) +\frac{1}{3} \int_x^0 \frac{d x'}{D(x',p)}
\exp\left[\int_{x'}^0 dx'' 
\frac{u(x'')}{D(x'',p)}\right] \frac{1}{p^2} \frac{\partial}{\partial p}
\int_{-\infty}^{x'} d x'' \frac{d u}{d x''} \left[f(x'',p) p^3\right]
\right\}.
\label{eq:complete}
\end{equation}
In the case of a spatially constant diffusion coefficient, as shown by 
\cite{malkov1}, a very good 
approximation to the solution is found in the form $f(x,p) =
f_0(p) \exp\left[-\frac{q(p)}{3D(p)}\int_x^0 dx'u(x')\right]$, with
$q(p) = -\frac{d\ln f_0}{d\ln p}$ and $f_0(p)=f(x=0,p)$ the
distribution function at the shock.
We found that the similar form 
\begin{equation}
f(x,p) = f_0(p) \exp\left[-\frac{q(p)}{3}\int_x^0 dx' \frac{u(x')}{D(x',p)}
\right]
\label{eq:solution}
\end{equation}
represents a very good approximation for the case of diffusion coefficients 
with arbitrary spatial dependence (see Sec.3). We adopt therefore this 
functional form in our 
calculations, although it is not strictly required, in the sense that
we could well use the complete solution, Eq. \ref{eq:complete}.

It was shown by \cite{blasi1} that the function $f_0(p)$ can be written
in general as
\begin{equation}
f_0 (p) = \left(\frac{3 R_{tot}}{R_{tot} U(p) - 1}\right) 
\frac{\eta n_0}{4\pi p_{inj}^3} 
\exp \left\{-\int_{p_{inj}}^p 
\frac{dp'}{p'} \frac{3R_{tot}U(p')}{R_{tot} U(p') - 1}\right\}.
\label{eq:inje}
\end{equation}
Here we introduced the function $U(p)=u_p/u_0$, with
\begin{equation}
u_p = u_1 - \frac{1}{f_0(p)} 
\int_{-\infty}^0 dx (du/dx)f(x,p)\ ,
\label{eq:up}
\end{equation}
where $u_1$ is the fluid velocity 
immediately upstream (at $x=0^-$).
We used $Q(x,p) = \frac{\eta n_{gas,1} u_1}{4\pi p_{inj}^2} 
\delta(p-p_{inj})\delta(x)$, with $n_{gas,1}=n_0 R_{tot}/R_{sub}$ the 
gas density immediately upstream ($x=0^-$) and $\eta$ the fraction of 
the particles crossing the shock which are going to take part in the 
acceleration process.
In Eq.~\ref{eq:inje} we also introduced the
two quantities $R_{sub}=u_1/u_2$ (compression factor at the subshock)
and $R_{tot}=u_0/u_2$ (total compression factor). The two compression 
factors, assuming, for simplicity, that the heating is only adiabatic, 
are related through the following expression (\cite{blasi1}):
\begin{equation}
R_{tot} = M_0^{\frac{2}{\gamma_g+1}} \left[ 
\frac{(\gamma_g+1)R_{sub}^{\gamma_g} - (\gamma_g-1)R_{sub}^{\gamma_g+1}}{2}
\right]^{\frac{1}{\gamma_g+1}},
\label{eq:Rsub_Rtot}
\end{equation}
where $M_0$ is the Mach number of the fluid at upstream infinity and 
$\gamma_g$ is the ratio of specific heats for the fluid. The parameter
$\eta$ in Eq.~\ref{eq:inje} contains the very important information 
about the injection of particles from the thermal bath. Following the 
work of \cite{vannoni}, we relate $\eta$ to the compression factor at 
the subshock as:
\begin{equation}
\eta = \frac{4}{3\pi^{1/2}} (R_{sub}-1) \xi^3 e^{-\xi^2}.
\end{equation}
Here $\xi$ is a parameter that identifies the injection 
momentum as a multiple of the momentum of the thermal particles in
the downstream section ($p_{inj}=\xi p_{th,2}$). This recipe is inspired
to the {\it thermal leakage model} originally presented by \cite{gjk00}
(see also previous work by \cite{eje81,ellison81,ee84}).
The parameter $\xi$
is supposed to contain the information about the microscopic 
structure of the shock. For collisionless shock waves the thickness
of the shock is expected to be of the order of the Larmor radius
of the thermal particles in the shock vicinity, which is not a
very well defined concept because of the violent fluctuations in
the electromagnetic fields within the shock. A simple argument 
can be used to infer that $\xi$ is of the order of $2-4$
(\cite{vannoni}). 
For the numerical calculations that follow we always use $\xi=3.5$, that
allows for only a fraction of order $10^{-4}$ of the particles
crossing the shock to be injected in the accelerator. 

Eq.~\ref{eq:momcons} for the conservation of the momentum flux, 
once normalized
to $\rho_0 u_0^2$, is easily transformed to
\begin{equation}
\xi_c (x) = 1 + \frac{1}{\gamma_g M_0^2} - U(x) - \frac{1}{\gamma_g M_0^2}
U(x)^{-\gamma_g},
\label{eq:normalized1}
\end{equation}
where $\xi_c (x) = P_{CR}(x)/\rho_0 u_0^2$ and $U(x)=u(x)/u_0$. In terms of 
the distribution function (Eq.~\ref{eq:solution}), we can also write:
\begin{equation}
\xi_c (x) = \frac{4\pi}{3\rho_0 u_0^2} \int_{p_{inj}}^{p_{max}} dp\ p^3
v(p) f_0(p) \exp\left[ -\int_x^0 dx' \frac{U(x')}{x_p(x',p)}
\right],
\label{eq:normalized2}
\end{equation}
where for simplicity we introduced $x_p(x,p)=\frac{3D(p,x)}{q(p) u_0}$.

By differentiating Eq.~\ref{eq:normalized2} with respect to $x$ we obtain
\begin{equation}
\frac{d\xi_c}{dx} = -\lambda(x) \xi_c(x) U(x),
\label{eq:differ}
\end{equation}
where
\begin{equation}
\lambda(x)=<1/x_p>_{\xi_c}=
\frac{\int_{p_{inj}}^{p_{max}} dp~p^3 \frac{1}{x_p(x,p)} v(p) f_0(p) 
\exp\left[ -\int_x^0 dx' \frac{U(x')}{x_p(x',p)}\right]}
{\int_{p_{inj}}^{p_{max}} 
dp~p^3 v(p) f_0(p) \exp\left[ -\int_x^0 dx' \frac{U(x')}{x_p(x',p)}\right]},
\label{eq:lambda}
\end{equation}
and $U(x)$ is expressed as a function of $\xi_c(x)$ through 
Eq.~\ref{eq:normalized1}. 

Finally, after integration by parts of Eq.~\ref{eq:up}, one is
able to express $U(p)$ in terms of an integration involving $U(x)$ alone:
\begin{equation}
U(p) = \int_{-\infty}^0 dx\ U(x)^2 \frac{1}{x_p(x,p)}
\exp\left[ -\int_x^0 dx' \frac{U(x')}{x_p(x',p)}\right]\ ,
\label{eq:up2}
\end{equation}
which allows to easily calculate $f_0(p)$ through Eq.~\ref{eq:inje}. 

Eqs.~\ref{eq:normalized1} and \ref{eq:differ}
can be solved by iteration in the following way: for a fixed value of the 
compression factor at the subshock, $R_{sub}$, the value of the
dimensionless velocity at the shock is calculated as $U(0)=R_{sub}/
R_{tot}$. The corresponding pressure in the form of accelerated
particles is given by Eq.~\ref{eq:normalized1} as 
$\xi_{c}(0) = 1 + \frac{1}{\gamma_g M_0^2} -\frac{R_{sub}}{R_{tot}}
- \frac{1}{\gamma_g M_0^2} \left(\frac{R_{sub}}{R_{tot}}\right)^{-\gamma_g}$.
This is used as a boundary condition for Eq.~\ref{eq:differ}, where
the functions $U(x)$ and $\lambda(x)$ (and therefore $f_0(p)$) on the 
right hand side at the $k^{th}$ step of iteration are taken as the
functions at the step $(k-1)$. In this way the solution of 
Eq.~\ref{eq:normalized1} at the step $k$ is simply
\begin{equation}
\xi_c^{(k)}(x) = \xi_c(0) \exp\left[-\int_x^0 d x' 
\lambda^{(k-1)}(x') U^{(k-1)}(x')\right],
\end{equation}
with the correct limits when $x\to 0$ and $x\to -\infty$. At each 
step of iteration the functions $U(x)$, $f_0(p)$, $\lambda(x)$ are
recalculated (through Eq.~\ref{eq:normalized1}, 
Eqs.~\ref{eq:up2} and \ref{eq:inje}, and Eq.~\ref{eq:lambda}, respectively), 
until convergence is reached. The solution of this set of equations, 
however, is also a solution of our physical problem only if the pressure
in the form of accelerated particles as given by Eq.~\ref{eq:normalized1} 
coincides with that calculated by using the final $f_0(p)$ in 
Eq.~\ref{eq:normalized2}. This occurs only for one specific
value of $R_{sub}$, which fully determines the solution of
our problem. 

\section{Results}

The computational method illustrated in the previous section is 
very fast and allows one to determine the solution (namely
the velocity and density profiles in the precursor, the density
of accelerated particles as a function of momentum and location
in the upstream fluid and all the thermodynamic quantities 
related to the gas) for an arbitrary choice of the diffusion 
coefficient and for any values of the environmental parameters
(Mach number, density, maximum momentum).

\begin{figure}
\resizebox{\hsize}{!}{
\includegraphics{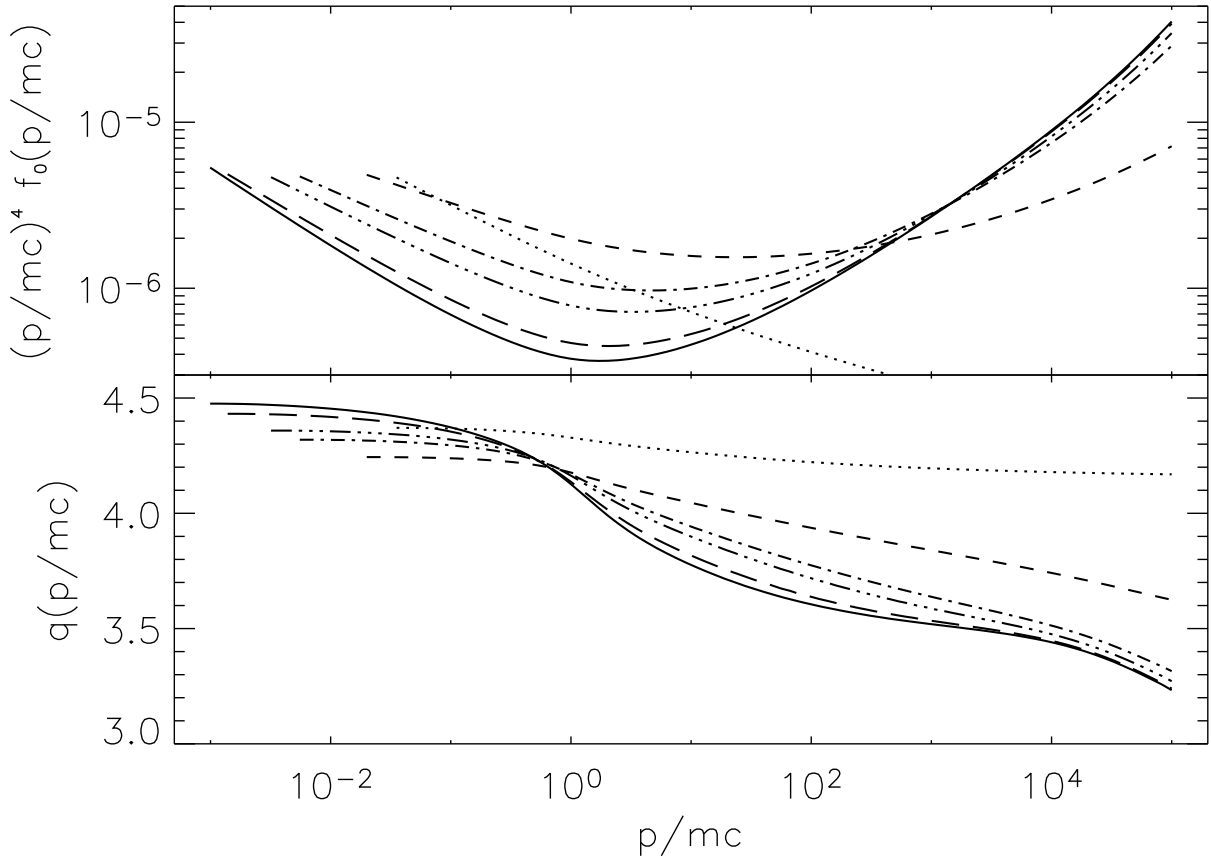}}
\caption{{\it Upper panel}: Spectra of accelerated particles at the 
location of the shock for 
$M_0=4$ (dotted line), 10 (short-dashed line), 50 (dash-dotted line), 100 
(dash-dot-dot-dotted line), 300 (long-dashed line) and 500 (solid line).
{\it Lower panel}: momentum dependent slope for the same values of 
Mach numbers. In both panels we used $\xi=3.5$ and $p_{max}=10^5 mc$.}
\label{fig:fig1}
\end{figure}
In Fig.~\ref{fig:fig1} we illustrate the spectra (upper panel) and slopes 
(lower panel) as a function of momentum for the following values
of the Mach number: $M_0=4$ (dotted line), 10 (short-dashed line), 
50 (dash-dotted line), 100 (dash-dot-dot-dotted line), 300 (long-dashed
line) and 500 (solid line). The distribution functions are multiplied by 
$p^4$ to emphasize the concave shape of the modified spectra. All the
curves refer to $p_{max}=10^5$ in units of $mc$. The most evident
aspect of shock modification, found in all previous calculations,
is here confirmed: the shock modification is enhanced when the
Mach number of the shock increases. The spectrum is flatter 
at high momenta as confirmed by the lower panel of Fig.~\ref{fig:fig1}, 
and easily understood in terms of the large values of the total
compression factor (see Table \ref{tbl-1}).

\begin{table*}
\begin{tabular}{cccccc}
\hline
Mach Number $M_0$ & $R_{sub}$ & $R_{tot}$ & $\xi_c(0)$ & $p_{inj}$ & $\eta$\\
\hline
$4$ & $3.19$ & $3.57$ & $0.1$ & $0.035$ &
$3.4\times 10^{-4}$  \\

$10$ & $3.413$ & $6.57$ & $0.47$ & $0.02$ &
$3.7\times 10^{-4}$  \\

$50$ & $3.27$ & $23.18$ & $0.85$ & $0.005$ &
$3.5\times 10^{-4}$  \\

$100$ & $3.21$ & $39.76$ & $0.91$ & $0.0032$ &
$3.4\times 10^{-4}$  \\

$300$ & $3.19$ & $91.06$ & $0.96$ & $0.0014$ &
$3.4\times 10^{-4}$  \\

$500$ & $3.29$ & $129.57$ & $0.97$ & $0.001$ &
$3.5\times 10^{-4}$  \\
\hline
\end{tabular}
\label{tbl-1}
\caption{Shock modification for different Mach numbers.}
\end{table*}

For strongly modified shocks, the slope becomes even flatter than 
$p^{-3.5}$ at high momenta, as also found in numerical simulations
(\cite{simple} and refs. therein) \footnote{We remind that in other 
semi-analytical 
calculations (e.g. \cite{malkov1}) the asymptotic spectrum for 
$p_{inj}\ll p < p_{max}$ has slope 3.5}. 
In these conditions, most energy is channeled in 
the highest energy part of the spectrum. At lower energies on the other
hand, the spectrum is steeper than that predicted by linear
theory, as a natural consequence of the lower compression at 
the subshock for strongly modified shocks. 
For the parameters adopted here, the energy saturation (namely
$\xi_c(0) \sim 1$) is achieved for Mach numbers around 100, as
demonstrated by the fact that the corresponding curves in the
upper panel of Fig.~\ref{fig:fig1} have roughly the same height (namely the
same energy content). On the other hand, different modifications
result in different compressions at the subshock and therefore
different injection momenta. This is illustrated in Fig.~\ref{fig:fig1} 
and Table~\ref{tbl-1}. In particular in Table~\ref{tbl-1} we list the
values of the compressione factors, dimensionless cosmic ray pressure at
the shock, injection momentum and fraction of accelerated particles for
the same values of $M_0$ used to obtain the curves in Fig.~\ref{fig:fig1}.

\begin{figure}
\resizebox{\hsize}{!}{
\includegraphics{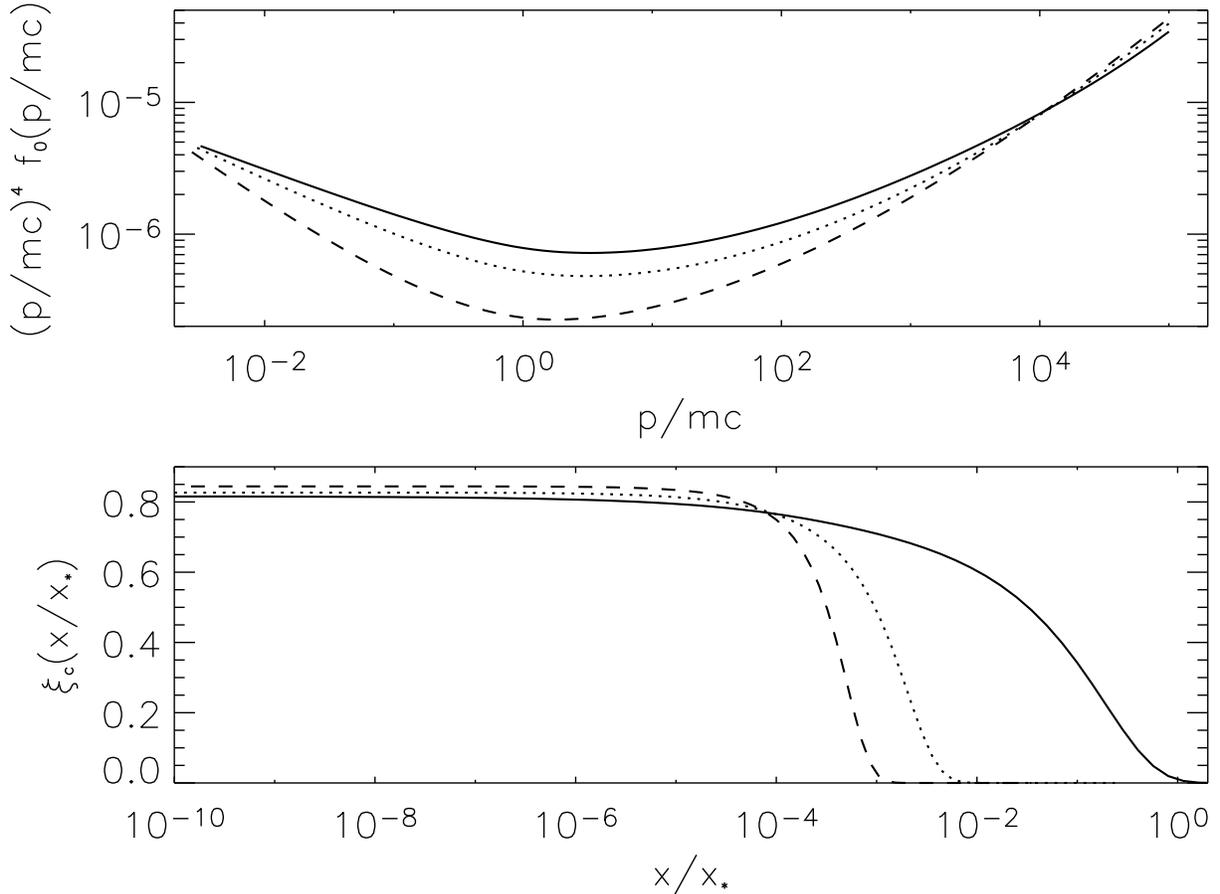}}
\caption{{\it Upper panel}: Spectra of accelerated particles at the
location of the shock for $M_0=100$, $p_{max}=10^5 mc$ and for a Bohm
diffusion coefficient (solid line), Kraichnan diffusion coefficient 
(dotted line) and Kolmogorov diffusion coefficient (dashed line). 
{\it Lower panel}: Distribution of the pressure in the form of
accelerated particles, normalized to the ram pressure ($\xi_c(x)$ as
defined by Eqs.~\ref{eq:normalized1}-\ref{eq:normalized2}), 
for the same three cases. The spatial coordinate is in units of 
$x_*=D_B(p_{\rm max})/u_0$, with $D_B$ the Bohm diffusion coefficient.}
\label{fig:fig2}
\end{figure}

In Fig.~\ref{fig:fig2} we illustrate the results of our method for different
choices of the momentum dependence of the diffusion coefficient. 
We consider three cases: 1) Bohm diffusion, $D_B(p)\propto p$;
2) Kraichnan diffusion, $D_{Kr}(p)\propto p^{1/2}$; 3) Kolmogorov
diffusion, $D_{Kol}(p)\propto p^{1/3}$ (relativistic scalings). 
For illustrative purposes, we
choose to calculate the spectrum of accelerated particles and
the shock modification for $M_0=100$ and $p_{max}=10^5~mc$. The 
resulting spectrum is shown in the upper panel of Fig.~\ref{fig:fig2}, for
Bohm diffusion (solid line), Kraichnan diffusion (dotted line)
and Kolmogorov diffusion (dashed line). The general tendency is
that the saturation phenomenon occurs at somewhat lower Mach numbers
for diffusion coefficients that depend more weakly on momentum.
The lower panel in Fig.~\ref{fig:fig2} illustrates the spatial distribution 
of energy in the accelerated particles ($\xi_c(x)$), where the
spatial coordinate is chosen in such a way that $x=1$ in the 
point $x_{*}=D_B(p_{max})/u_0$. Clearly the particles
with the maximum momentum diffuse on shorter spatial scales than $x_*$ 
for diffusion coefficients with weaker momentum dependence. 

\begin{figure}
\resizebox{\hsize}{!}{
\includegraphics{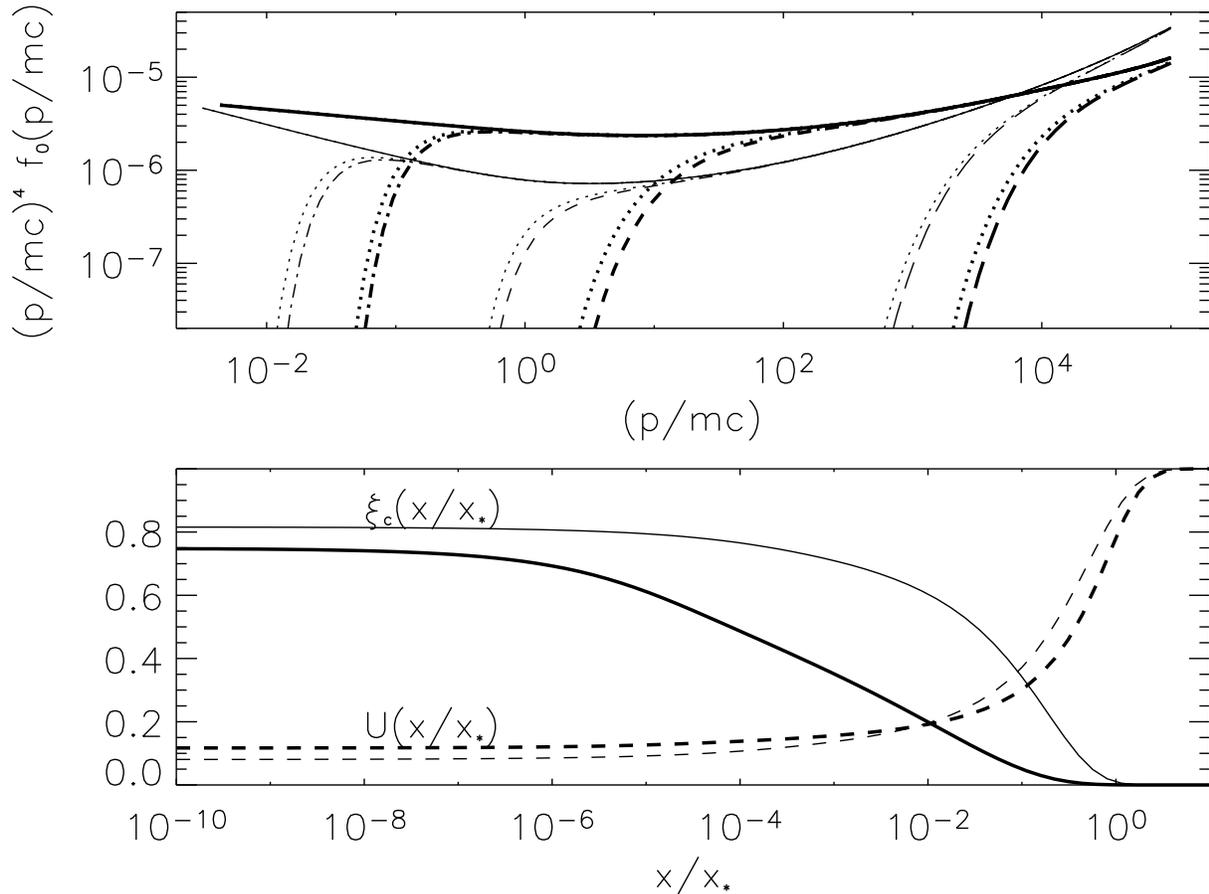}}
\caption{{\it Upper panel}: Spectra of the accelerated particles
for spatially constant Bohm diffusion (thin lines) and for Bohm 
diffusion with $D(p,x)\propto p/ \rho(x)$. The different line-types
refer, for each case, to $x=0$ (solid line), 
$x=10^{-7}\ x_*$ (dot-dashed line),
$x=10^{-4}\ x_*$ (short-dashed line) and $x=0.1\ x_*$ (long-dashed line).
The dotted lines neighbouring each curve refer to the distribution 
functions computed by using in the right-hand-side of 
Eq.~\ref{eq:complete} the solution
obtained with our method (see text for details).
{\it Lower panel}: $\xi_c(x)$ and $U(x)$ (solid and dashed lines, 
respectively) for the case of spatially constant Bohm diffusion 
(thin lines) and for $D(p,x)\propto p/ \rho(x)$ (thick lines). The spatial
coordinate is again in units of $x_*$ defined as for Fig.~\ref{fig:fig2}.
}
\label{fig:fig3}
\end{figure}
The power of the computational method in being suitable for treating
arbitrary dependences of the diffusion coefficient on momentum and 
spatial coordinates is further demonstrated in Fig.~\ref{fig:fig3}, 
where we show 
how the solutions change when the diffusion coefficient is allowed to 
vary in space. For illustrative purposes we consider the case of a 
Bohm diffusion coefficient with $D_B(p,x)\propto p$ (constant in space) 
and $D_B(p,x)\propto p/ \rho(x)$, where $\rho(x)$ is the gas density at 
the position $x$, self-consistently calculated by using the
conservation laws. The latter dependence is representative of the case
of a magnetic 
field frozen in the plasma flowing in the upstream section. 

In the upper panel of Fig.~\ref{fig:fig3} we plot the spectrum of 
the accelerated 
particles for spatially constant Bohm diffusion (thin curves) and
for $D_B(p,x)\propto p/ \rho(x)$ (thick lines). The different line-types
refer to spectra at the different spatial locations: 
$x=0$ (solid line), $x=10^{-7}x_*$ (dot-dashed line), 
$x=10^{-4}x_*$ (short-dashed line) and 
$x=0.1x_*$ (long-dashed line), where $x_*$ is defined as above, 
i.e. $x_*=D_B(p_{max})/u_0$ with $D_B$
referring to the spatially constant Bohm diffusion coefficient.
In the lower panel we plot 
$\xi_c(x)$ (solid lines) and $U(x)$ (dashed lines) for the same two
cases, identified by the different thickness of the lines. 

In order to assess the goodness of our approximate solution 
(Eq.~\ref{eq:solution}) we computed the right-hand-side of 
Eq.~\ref{eq:complete} by using the functions $U(x)$ and $f(x,p)$
found with our method. The correction is found to be non-negligible
only in the exponentially decreasing parts of the spectrum (see
dotted lines in Fig.~\ref{fig:fig3}), which contains negligible
energy and hardly leads to any observable features. On this
basis, we conclude that Eq.~\ref{eq:solution} is an excellent 
approximation to the solution for diffusion coefficients with arbitrary
spatial and momentum dependences.

The solutions obtained with this method are remarkably similar to
those obtained with approximate methods by \cite{blasi1,blasi2}, for
the case of Bohm diffusion with no spatial dependence. The
discrepancies with such previous treatments are expected and 
indeed appear for increasingly weaker dependences of the diffusion 
coefficient on the particles' momentum, and in general when a 
spatial dependence 
of the diffusion properties is added (these aspects will be
discussed in an upcoming detailed paper). The results also compare
well with a previous method proposed by \cite{malkov1} and
\cite{malkov2}, 
where the contribution of gas pressure was neglected and no recipe for 
injection was adopted. 

The most important property of the method here described, however, is
the fact that it appears to be the
first that allows to take into account the spatial dependence of
the diffusion coefficient. The importance of being able to deal with
arbitrary diffusion properties is highlighted by the following
considerations. First, particle acceleration at shocks is 
expected to be efficient only if the turbulence responsible for 
diffusion is self-generated (\cite{lc83,lb00,bell04}), and 
in this case the diffusion
coefficient is necessarily dependent upon both momentum and space in a 
complex manner. Moreover, the appearance of a maximum momentum is indeed 
due to the fact that at some distance from the shock diffusion becomes
uneffective and particles are no longer trapped in the shock
vicinity.
Since the shock modification depends in a crucial way on the value of
the maximum momentum, it is clear that a careful calculation of the 
shock modification should be able to account for these phenomena.

\section*{Acknowledgments}
This research was funded through grant COFIN2004. We wish to
acknowledge useful conversations with D. Ellison, S. Gabici and M. Vietri.

\end{document}